\documentclass[pra,twocolumn,showpacs,showkeys,amsmath,amssymb]{revtex4}
\usepackage{graphicx}   
\usepackage{hyperref}
\usepackage{bm}     
\begin{document}


\title{The non-Markovian quantum behavior of open systems:
       An exact Monte Carlo method employing stochastic product states}

\author{Heinz-Peter Breuer}
\email{breuer@theorie.physik.uni-oldenburg.de}
\affiliation{Institut f\"ur Physik, Carl von Ossietzky
             Universit\"at, D-26111 Oldenburg, Germany}
\affiliation{Physikalisches Institut, Universit\"at Freiburg,
             D-79104 Freiburg, Germany}
\date{\today}
\begin{abstract}
It is shown that the exact dynamics of a composite quantum system
can be represented through a pair of product states which evolve
according to a Markovian random jump process. This representation
is used to design a general Monte Carlo wave function method that
enables the stochastic treatment of the full non-Markovian
behavior of open quantum systems. Numerical simulations are
carried out which demonstrate that the method is applicable to
open systems strongly coupled to a bosonic reservoir, as well as
to the interaction with a spin bath. Full details of the
simulation algorithms are given, together with an investigation of
the dynamics of fluctuations. Several potential generalizations of
the method are outlined.
\end{abstract}
\maketitle

\section{Introduction}
A great deal of the dynamics of open systems can be described, to
a reasonable degree of accuracy, by Markovian quantum master
equations. Important examples are given by the weak-coupling
interaction of radiation with matter in atomic physics and quantum
optics \cite{COHEN,GARDINER}. However, non-Markovian quantum
dynamics \cite{NAKAJIMA,ZWANZIG,FEYNMAN,LEGGETT} is known to play
a significant role in many applications of the theory of open
quantum systems \cite{TheWork} currently under discussion in the
literature, e.g. the dynamics of the atom laser \cite{SAVAGE},
environment-induced decoherence at low temperatures (for an
example, see \cite{DECOHERENCE}), and quantum devices interacting
with a spin bath \cite{STAMP}.

Quantum Monte Carlo techniques have been shown to provide
efficient numerical tools for the treatment of the dynamics of
open systems in the Markovian regime \cite{SWFM}. In these
techniques one constructs a stochastic dynamics for the open
system's state vector $\psi(t)$ such that the reduced density
matrix $\rho_S(t)$ of the open system is recovered through the
expression
$\rho_S(t)={\mathrm{E}}(|\psi(t)\rangle\langle\psi(t)|)$, where
${\mathrm{E}}$ denotes the expectation value or ensemble average
of the underlying process. This is the standard Monte Carlo wave
function method which has been widely used in many physical
problems of quantum optics and condensed matter theory.

The idea of the Monte Carlo wave function method can be extended
to the treatment of non-Markovian quantum processes which cannot
be described by a Markovian quantum master equation. One such
method \cite{DGS} is based on a stochastic integro-differential
equation for the wave function involving a non-local retarded
memory kernel. The solution of non-local equations of motion can
be circumvented by employing a {\textit{pair}} $\psi_1(t)$,
$\psi_2(t)$ of random wave functions of the open system and by
expressing the reduced density matrix with the help of the mean
value $\rho_S(t)={\mathrm{E}}(|\psi_1(t)\rangle\langle\psi_2(t)|)$
\cite{BKP}. This method of propagating a pair of wave functions
requires the construction of an appropriate time-local
non-Markovian master equation. Such an equation can be obtained
with the help of the time-convolutionless (TCL) projection
operator technique which leads to a systematic perturbation
expansion for the time-dependent generator of the master equation.
However, for strong system-environment couplings calculations
based on the TCL expansion become extremely complicated and the
derivation of an appropriate TCL generator of high order is, in
general, not feasible in practice. A further possibility is to use
an explicit expression for the influence functional of the open
system to obtain stochastic differential equations for a pair of
random wave functions \cite{STOCK}. This method is, however,
restricted to Gaussian reservoirs and linear dissipation.

In this paper the details of a new method proposed in \cite{PDP-SHORT}
are presented, which allows to attack the
problem of non-Markovian quantum evolution by means of a Monte
Carlo wave function technique. The basic idea is to introduce a
pair $|\Phi_1(t)\rangle$, $|\Phi_2(t)\rangle$ of random states of
the {\textit{total}} system, with the aim of a stochastic
formulation of the exact von Neumann dynamics of the composite
system. A similar idea has been used recently to
construct an exact diffusion process for a pair of one-particle
wave functions describing systems of identical Bosons
\cite{CARUSO1} and Fermions \cite{CHOMAZ}. Here, the state vector
dynamics is assumed to represent a piecewise deterministic process
(PDP). This is a Markovian jump process with smooth,
deterministic evolution periods between successive jumps. The
stochastic states of the total system are supposed to be tensor
product states of the form $|\Phi_1\rangle=\psi_1\otimes\chi_1$
and $|\Phi_2\rangle=\psi_2\otimes\chi_2$. The method thus involves
four stochastic state vectors, namely a pair $\psi_1$, $\psi_2$ of
state vectors of the open system, and a pair $\chi_1$, $\chi_2$ of
state vectors of the environment. The open system's reduced
density matrix can then be represented in terms of the expectation
value
\begin{equation} \label{RHOS}
 \rho_S(t) = {\mathrm{E}} \left(
 |\psi_1(t)\rangle\langle\psi_2(t)|
 \langle \chi_2(t) | \chi_1(t) \rangle \right).
\end{equation}
Contrary to the standard methods mentioned above, this
representation employs an average over the product of two
quantities: The dyadic $|\psi_1\rangle\langle\psi_2|$ of a pair of
state vectors of the open system, and the scalar product $\langle
\chi_2 | \chi_1 \rangle$ of a corresponding pair of environment
states. It will be shown that this representation allows to design
a Markovian stochastic process which unravels the full
non-Markovian behavior of the reduced density matrix.

The paper is structured as follows. Section II contains the
general construction of the PDP representing the exact von Neumann
dynamics of the composite system, an investigation of the dynamics
of the fluctuations of the stochastic process, as well as a
detailed description of the Monte Carlo algorithm of the open
system dynamics. The example of the non-perturbative decay of a
two-state system into a bosonic reservoir is discussed in
Sec.~III. This section contains numerical simulations of the
non-Markovian dynamics of the decay into a reservoir in the regime 
of strong couplings and corresponding long memory times. The quantum
dynamics of a specific spin bath model is investigated in Sec.~IV.
This model describes the interaction of a single electron spin in
a quantum dot with an external magnetic field and a bath of
nuclear spins. Section V contains the conclusions and indicates
various potential generalizations of the stochastic method.

\section{General formulation of the method} \label{GENERAL}

\subsection{Construction of the PDP} \label{PDP}
We investigate the general situation of an open system with
underlying Hilbert space ${\mathcal{H}}_S$, which is coupled to an
environment with Hilbert space ${\mathcal{H}}_E$. The state space
of the composite, total quantum system is given by the tensor
product ${\mathcal{H}}_S\otimes{\mathcal{H}}_E$. Working in the
interaction picture we write the Hamiltonian describing the
system-environment interaction as
\begin{equation}
 H_I(t) = \sum_{\alpha} A_{\alpha}(t) \otimes B_{\alpha}(t).
\end{equation}
The $A_{\alpha}(t)$ and the $B_{\alpha}(t)$ are interaction
picture operators acting in ${\mathcal{H}}_S$ and
${\mathcal{H}}_E$, respectively. The evolution of the density
matrix $\rho(t)$ of the total system is then governed by the von
Neumann equation ($\hbar = 1$),
\begin{equation} \label{NEUMANN}
 \frac{d}{dt} \rho(t) = -i [H_I(t),\rho(t)].
\end{equation}
Our central goal is to construct a representation of $\rho(t)$ in
terms of the expectation value
\begin{equation} \label{DEFPHINU}
 \rho(t) = {\rm{E}}(|\Phi_1(t)\rangle\langle\Phi_2(t)|),
\end{equation}
which is determined through a pair $|\Phi_1(t)\rangle$,
$|\Phi_2(t)\rangle$ of stochastic state vectors of the composite
quantum system. Equivalently, one may define the quantity
\begin{equation} \label{DEFR}
 R(t) = |\Phi_1(t)\rangle\langle\Phi_2(t)|,
\end{equation}
which is a random operator on
${\mathcal{H}}_S\otimes{\mathcal{H}}_E$, and write the density
matrix as the mean value of this operator, that is
$\rho(t)={\rm{E}}(R(t))$.

In the following we suppose that the stochastic state vectors
$|\Phi_\nu(t)\rangle$ ($\nu=1,2$) introduced in
Eq.~(\ref{DEFPHINU}) are direct products of certain system states
$\psi_{\nu}(t) \in {\mathcal{H}}_S$ and environment states
$\chi_{\nu}(t) \in {\mathcal{H}}_E$, that is we have
\begin{equation} \label{DIRECT-PRODUCT}
 |\Phi_\nu(t)\rangle = \psi_{\nu}(t) \otimes \chi_{\nu}(t),
 \qquad \nu = 1,2.
\end{equation}
The reduced density matrix $\rho_S(t)$ of the open system is
defined through the partial trace over the variables of the
environment, $\rho_S(t)={\mathrm{tr}}_E\rho(t)$. In view of
Eqs.~(\ref{DEFPHINU}) and (\ref{DIRECT-PRODUCT}) this definition
immediately leads to the relation (\ref{RHOS}).

It is important to realize that a representation of the form given
in Eqs.~(\ref{DEFPHINU}) and (\ref{DIRECT-PRODUCT}) is possible
for any initial state $\rho(t=0)$. This means that any given
density matrix $\rho\equiv\rho(0)$ of the composite quantum system
can be written as the mean value
$\rho={\rm{E}}(|\Phi_1\rangle\langle\Phi_2|)$, in which the random
states $|\Phi_\nu\rangle$ are direct products of the form
(\ref{DIRECT-PRODUCT}). In particular, it is {\textit{not}}
necessary to demand that $\rho$ describes an initial state without
system-environment correlations.

A formal proof of this statement may be carried out as follows.
One first observes that a sequence of pairs
$(|\Phi_1^{\lambda}\rangle,|\Phi_2^{\lambda}\rangle)$ of state
vectors, which occur with corresponding probabilities
$p_{\lambda}$, gives rise to the expectation value
\begin{equation} \label{PROOF1}
 \rho = {\rm{E}}(|\Phi_1\rangle\langle\Phi_2|)
 = \sum_{\lambda} p_{\lambda}
 |\Phi^{\lambda}_1\rangle\langle\Phi^{\lambda}_2|.
\end{equation}
Of course, $p_{\lambda}$ provides a probability distribution
satisfying $p_{\lambda}\geq 0$ and $\sum_{\lambda}p_{\lambda}=1$.
Introducing new states through the relation
$|\Psi_{\nu}^\lambda\rangle=\sqrt{p_{\lambda}}|\Phi_{\nu}^{\lambda}\rangle$,
we can write
\begin{equation} \label{PROOF2}
 \rho = \sum_{\lambda}
 |\Psi^{\lambda}_1\rangle\langle\Psi^{\lambda}_2|.
\end{equation}
Thus, to prove the above statement we have to show that any given
density matrix $\rho$ of the composite quantum system can be
brought into the form (\ref{PROOF2}), whereby the
$|\Psi^{\lambda}_{\nu}\rangle$ must be direct products. To
demonstrate that this is in fact possible we introduce an ON-basis
$\{\psi_i\}$ in ${\mathcal{H}}_S$ and an ON-basis $\{\chi_n\}$ in
${\mathcal{H}}_E$ and write the given $\rho$ as follows,
\begin{equation} \label{PROOF3}
 \rho = \sum_{ijnm} \rho_{ijnm}
 |\psi_i\rangle\langle\psi_j| \otimes
 |\chi_n\rangle\langle\chi_m|,
\end{equation}
where
\[
 \rho_{ijnm} \equiv \langle \psi_i\chi_n|\rho|\psi_j\chi_m\rangle
 \equiv |\rho_{ijnm}|e^{2i\varphi_{ijnm}}.
\]
Next, one introduces a collective index $\lambda=(ijnm)$ and
defines the states
\begin{eqnarray}
 |\Psi^{\lambda}_1\rangle &=& \sqrt{|\rho_{ijnm}|}e^{+i\varphi_{ijnm}}
 \psi_i \otimes \chi_n, \label{PROOF5} \\
 |\Psi^{\lambda}_2\rangle &=& \sqrt{|\rho_{ijnm}|}e^{-i\varphi_{ijnm}}
 \psi_j \otimes \chi_m, \label{PROOF6}
\end{eqnarray}
which allow one to write Eq.~(\ref{PROOF3}) in the desired form
(\ref{PROOF2}). This completes the proof since the states
(\ref{PROOF5}) and (\ref{PROOF6}) are indeed direct products.

The aim is now to construct an appropriate stochastic process for
the state vectors $|\Phi_{\nu}(t)\rangle$ which exactly reproduces
the von Neumann equation (\ref{NEUMANN}) through the expectation
value (\ref{DEFPHINU}). As mentioned in the Introduction we
suppose that the time-evolution represents a piecewise
deterministic process (PDP). A convenient way of formulating a PDP
is to write stochastic differential equations for the random
variables. The foundations of the calculus of PDPs and its
applications to the quantum theory of open systems may be found in
\cite{TheWork}. In view of the representation
(\ref{DIRECT-PRODUCT}) the stochastic dynamics can be defined in
terms of stochastic differential equations for the state vectors
$\psi_{\nu}(t)$ and $\chi_{\nu}(t)$,
\begin{eqnarray}
 d\psi_{\nu}(t) &=& F_{\nu}dt + dJ_{\nu}, \label{STOCHDGL1} \\
 d\chi_{\nu}(t) &=& G_{\nu}dt + dK_{\nu}. \label{STOCHDGL2}
\end{eqnarray}
These equations reflect the general structure of a PDP: The terms
$F_{\nu}dt$ and $G_{\nu}dt$ represent the deterministic evolution
periods, the drift of the process, while the terms $dJ_{\nu}$ and
$dK_{\nu}$ provide the contributions from the random,
instantaneous jumps of the process. These jump contributions are
taken to be of the form
\begin{eqnarray}
 dJ_{\nu} &=& \sum_{\alpha}
 \left( -iL_{\alpha\nu}A_{\alpha}-I\right) \psi_{\nu} dN_{\alpha\nu}(t),
 \label{JUMPS1} \\
 dK_{\nu} &=& \sum_{\alpha}
 \left( M_{\alpha\nu}B_{\alpha}-I\right) \chi_{\nu}
 dN_{\alpha\nu}(t).
 \label{JUMPS2}
\end{eqnarray}
Here, $I$ denotes the identity operator and $L_{\alpha\nu}$,
$M_{\alpha\nu}$ are c-number functionals which will be specified
below. The quantities $dN_{\alpha\nu}(t)$ are known as Poisson
increments. They are independent, random numbers which take on the
possible values $0$ or $1$ and satisfy the relation
\begin{equation} \label{DNDN}
 dN_{\alpha\nu}(t) dN_{\beta\mu}(t) =
 \delta_{\alpha\beta} \delta_{\nu\mu} dN_{\alpha\nu}(t).
\end{equation}
Under the condition that $dN_{\alpha\nu}(t)=1$ for a particular
$\alpha$ and $\nu$ the other Poisson increments therefore vanish
and, by virtue of the Eqs.~(\ref{JUMPS1}) and (\ref{JUMPS2}), the
state vectors then carry out the instantaneous jumps
\begin{equation} \label{JUMPS}
 \psi_{\nu} \longrightarrow -iL_{\alpha\nu}A_{\alpha}\psi_{\nu},
 \qquad
 \chi_{\nu} \longrightarrow M_{\alpha\nu}B_{\alpha}\chi_{\nu}.
\end{equation}
The expectation values of the Poisson increments are given by
\begin{equation} \label{EDN}
 {\mathrm{E}}(dN_{\alpha\nu}(t)) = \Gamma_{\alpha\nu}dt.
\end{equation}
This implies that $dN_{\alpha\nu}(t)=1$  with probability
$\Gamma_{\alpha\nu}dt$ and, hence, the jumps (\ref{JUMPS}) occur
at a rate $\Gamma_{\alpha\nu}$, which will also be determined
below. If, on the other hand, all Poisson increments vanish we
have $d\psi_{\nu}(t)=F_{\nu}dt$ and $d\chi_{\nu}(t)=G_{\nu}dt$,
which means that the state vectors follow the deterministic drift
during $dt$.

Our next step consists in deriving a stochastic equation for the
random operator $R(t)$ defined in Eq.~(\ref{DEFR}), which will
then lead to an equation of motion for the expectation value
(\ref{DEFPHINU}). Employing the calculus of PDPs one finds
\[
 dR = |d\Phi_1\rangle\langle\Phi_2| + |\Phi_1\rangle\langle d\Phi_2|
      + |d\Phi_1\rangle\langle d\Phi_2|.
\]
The third term on the right-hand side of this equation involves
the products $dN_{\alpha 1}dN_{\beta 2}$ of the Poisson
increments, which vanish by virtue of Eq.~(\ref{DNDN}). This means
that the state vectors $|\Phi_1(t)\rangle$ and $|\Phi_2(t)\rangle$
evolve independently and that we may write
\begin{equation} \label{DRT}
 dR=|d\Phi_1\rangle\langle\Phi_2|+|\Phi_1\rangle\langle d\Phi_2|.
\end{equation}
With the help of the stochastic differential equations
(\ref{STOCHDGL1}) and (\ref{STOCHDGL2}) the state vector
increments are found to be
\begin{eqnarray*}
 |d\Phi_{\nu}\rangle &=& d\psi_{\nu} \otimes \chi_{\nu}
 + \psi_{\nu} \otimes d\chi_{\nu} + d\psi_{\nu} \otimes
 d\chi_{\nu} \\
 &=& (F_{\nu}dt + dJ_{\nu}) \otimes \chi_{\nu}
     + \psi_{\nu} \otimes (G_{\nu}dt + dK_{\nu}) \\
 &~& + dJ_{\nu} \otimes dK_{\nu}.
\end{eqnarray*}
On using the structure of the jump terms (\ref{JUMPS1}) and
(\ref{JUMPS2}) and relation (\ref{DNDN}) the third term may be
written
\begin{eqnarray*}
 dJ_{\nu} \otimes dK_{\nu} &=&
 \sum_{\alpha} \left(-iL_{\alpha\nu}A_{\alpha}-I\right)\psi_{\nu}
 \\ &~& \qquad \otimes
 \left(M_{\alpha\nu}B_{\alpha}-I\right)\chi_{\nu}dN_{\alpha\nu} \\
 &=& -dJ_{\nu} \otimes \chi_{\nu} \\
 &~& + \sum_{\alpha} \left(-iL_{\alpha\nu}A_{\alpha}-I\right)\psi_{\nu}
 \\ &~& \qquad \otimes
 M_{\alpha\nu}B_{\alpha}\chi_{\nu}dN_{\alpha\nu},
\end{eqnarray*}
which leads to
\begin{eqnarray}
 |d\Phi_{\nu}\rangle
 &=& F_{\nu}dt \otimes \chi_{\nu} \\
 &~& + \psi_{\nu} \otimes \left(G_{\nu}dt - \sum_{\alpha}
     dN_{\alpha\nu} \chi_{\nu}\right) \nonumber \\
 &~& -i\sum_{\alpha} L_{\alpha\nu}M_{\alpha\nu} (A_{\alpha}\psi_{\nu})
 \otimes (B_{\alpha}\chi_{\nu}) dN_{\alpha\nu}. \nonumber
\end{eqnarray}
This equation provides an exact relation for the stochastic
increments $|d\Phi_{\nu}\rangle$. To ensure that the first and the
second term on the right-hand side vanish when taking the average
over the Poisson increments, we now set
\begin{equation}
 F_{\nu} \equiv 0, \qquad G_{\nu} \equiv \Gamma_{\nu} \chi_{\nu},
\end{equation}
where
\begin{equation} \label{DEFGNU}
 \Gamma_{\nu} \equiv \sum_{\alpha} \Gamma_{\alpha\nu},
\end{equation}
and
\begin{equation} \label{DEFGAN}
 \Gamma_{\alpha\nu} \equiv \frac{1}{L_{\alpha\nu}M_{\alpha\nu}}.
\end{equation}
This yields the expression
\begin{eqnarray} \label{DPHINU}
 |d\Phi_{\nu}\rangle &=&
 \psi_{\nu} \otimes \left(\Gamma_{\nu}dt
 - \sum_{\alpha} dN_{\alpha\nu}\right) \chi_{\nu} \\
 &~& -i\sum_{\alpha}  \Gamma^{-1}_{\alpha\nu} (A_{\alpha}\psi_{\nu})
 \otimes (B_{\alpha}\chi_{\nu}) dN_{\alpha\nu}. \nonumber
\end{eqnarray}
Finally, we substitute (\ref{DPHINU}) into (\ref{DRT}) to arrive
at
\begin{equation} \label{DR}
 dR(t) = -i [H_I(t),R(t)]dt + dS(t).
\end{equation}

Equation (\ref{DR}) is the desired exact stochastic equation of
motion of the random operator $R(t)$. The drift term involves the
commutator with the interaction Hamiltonian $H_I(t)$, while the
noise term is given by the stochastic increment
\begin{equation} \label{DS}
 dS(t) = dT_1 R(t) + R(t) dT_2^{\dagger},
\end{equation}
with
\begin{equation} \label{DT}
 dT_{\nu} = \sum_{\alpha}
 \left( \Gamma_{\alpha\nu} dt - dN_{\alpha\nu} \right)
 \left(I+i\Gamma^{-1}_{\alpha\nu} A_{\alpha}B_{\alpha} \right).
\end{equation}
According to Eqs.~(\ref{EDN}) and (\ref{DT}) the average over the
Poisson increments yields ${\mathrm{E}}(dT_{\nu})=0$. By virtue of
Eq.~(\ref{DS}) this gives ${\mathrm{E}}(dS)=0$. Thus, if we take
the average of both sides of Eq.~(\ref{DR}) we are led directly to
the von Neumann equation (\ref{NEUMANN}). This shows that on
average the stochastic dynamics defined by the differential
equations (\ref{STOCHDGL1}) and (\ref{STOCHDGL2}) indeed
reproduces the exact von Neumann dynamics of the density matrix of
the combined system. We have thus achieved the goal of
constructing a stochastic formulation of the evolution of the
total system by means of a Markovian piecewise deterministic
process.

Up to this point the quantities $L_{\alpha\nu}$ and
$M_{\alpha\nu}$ are completely arbitrary with the only restriction
that $\Gamma_{\alpha\nu}\geq 0$ (see Eq.~(\ref{DEFGAN})), which
guarantees that the expectation values
${\mathrm{E}}(dN_{\alpha\nu})$ are positive, as it should be for
random Poisson increments (see Eq.~(\ref{EDN})). In the following
we choose
\begin{equation} \label{DEFLM}
 L_{\alpha\nu} = \frac{||\psi_{\nu}||}{||A_{\alpha}\psi_{\nu}||},
 \qquad
 M_{\alpha\nu} = \frac{||\chi_{\nu}||}{||B_{\alpha}\chi_{\nu}||}.
\end{equation}
The advantage of this choice is that the jumps described by
Eq.~(\ref{JUMPS}) then conserve the norm of the stochastic state
vectors $\psi_{\nu}$ and $\chi_{\nu}$.
Summarizing, the stochastic differential equations defining the
PDP now read as follows,
\begin{eqnarray}
 d\psi_{\nu} &=& \sum_{\alpha}
 \left( \frac{-i||\psi_{\nu}||}{||A_{\alpha}\psi_{\nu}||}A_{\alpha}-I\right)
 \psi_{\nu} dN_{\alpha\nu}(t), \label{STOCH1} \\
 d\chi_{\nu} &=& \Gamma_{\nu} \chi_{\nu} dt \nonumber \\
 &~& + \sum_{\alpha}
 \left( \frac{||\chi_{\nu}||}{||B_{\alpha}\chi_{\nu}||}B_{\alpha}-I\right)
 \chi_{\nu} dN_{\alpha\nu}(t), \label{STOCH2}
\end{eqnarray}
where $\Gamma_{\nu}$ is given by Eq.~(\ref{DEFGNU}) and by
\begin{equation} \label{DEFGANU}
 \Gamma_{\alpha\nu} =
 \frac{||A_{\alpha}\psi_{\nu}||\cdot||B_{\alpha}\chi_{\nu}||}
 {||\psi_{\nu}||\cdot||\chi_{\nu}||}.
\end{equation}
We observe that $\psi_{\nu}(t)$ is a pure, norm-conserving jump
process, while $\chi_{\nu}(t)$ is a PDP with norm-conserving jumps
and a linear drift which leads to a monotonic increase of the norm
of $\chi_{\nu}$.

\subsection{Dynamics of fluctuations} \label{FLUCTUATIONS}
As a measure of the size of the fluctuations of the stochastic
process constructed above we define \cite{CARUSO1}
\begin{eqnarray} \label{D2t}
 D^2(t) &\equiv& {\mathrm{E}}\left( ||R(t)-\rho(t) ||^2 \right)
 \nonumber \\
 &=& {\mathrm{E}}\left( {\mathrm{tr}}\left\{ \left[
 R(t)-\rho(t) \right]^{\dagger} \left[ R(t) - \rho(t) \right] \right\}
 \right).
\end{eqnarray}
The quantity $D(t)$ is thus the root mean square distance from the
stochastic operator $R(t)$ to its mean value
$\rho(t)={\mathrm{E}}(R(t))$, the distance being determined
through the Hilbert-Schmidt norm
$||A||=\sqrt{\mathrm{tr}\{A^{\dagger}A\}}$, where the trace is
taken over the Hilbert space of the total system. Equation
(\ref{D2t}) may be written as
\begin{equation} \label{D2t2}
 D^2(t) = {\mathrm{E}}\left( {\mathrm{tr}}\left\{R^{\dagger}(t)R(t)\right\}
 \right) - {\mathrm{tr}} \rho^2(t).
\end{equation}
Since the dynamics of $\rho(t)$ represents a unitary
transformation the trace over the square of $\rho(t)$ is constant
in time. For a pure initial state $\rho(0)$ we have ${\mathrm{tr}}
\rho^2 \equiv 1$. Moreover, in the case of a sharp initial state,
that is for $R(0)=\rho(0)$, one finds that $D^2(0) = 0$.

Our aim is to estimate the size of the fluctuations. To this end
we first derive a differential equation for the mean square
distance $D^2(t)$. With the help of the stochastic equation of
motion (\ref{DR}) and of definition (\ref{DS}) the differential of
$D^2(t)$ is found to be
\begin{eqnarray} \label{D2t3}
 dD^2 &=& {\mathrm{E}}
 \left({\mathrm{tr}}\left\{dS^{\dagger}dS\right\}\right)
 \nonumber \\
 &=& {\mathrm{E}}\left({\mathrm{tr}}\left\{
 dT_1^{\dagger}dT_1 RR^{\dagger}+dT_2^{\dagger}dT_2
 R^{\dagger}R\right\}\right).
\end{eqnarray}
Using then the definition (\ref{DT}) of the quantities $dT_{\nu}$
as well as Eqs.~(\ref{DEFR}), (\ref{DNDN}) and (\ref{EDN}), we
obtain
\[
 \frac{dD^2}{dt} = {\mathrm{E}}\left(
 \sum_{\alpha\nu} \Gamma_{\alpha\nu} \frac{||
 \left(I+i\Gamma_{\alpha\nu}^{-1}A_{\alpha}B_{\alpha}\right)
 |\Phi_{\nu}\rangle||^2}{||\,|\Phi_{\nu}\rangle||^2}
 {\mathrm{tr}}\left\{R^{\dagger}R\right\}\right) \!\! .
\]
The choice (\ref{DEFLM}) finally yields
\begin{equation} \label{D2t5}
 \frac{dD^2}{dt} = 2 {\mathrm{E}}\left(
 \sum_{\nu} \Gamma_{\nu}
 {\mathrm{tr}}\left\{R^{\dagger}R\right\}\right).
\end{equation}

Equation (\ref{D2t5}) is an exact differential equation for the
fluctuations of the random process. To find a rough estimate of
the size of the fluctuations we suppose that the rates
$\Gamma_{\nu}$ are bounded from above, that is $\Gamma_{\nu} \leq
\Gamma_0$. This leads to the inequality
\begin{equation} \label{D2t6}
 \frac{dD^2}{dt} \leq 4 \Gamma_0
 \left( D^2 + {\mathrm{tr}}\rho^2 \right),
\end{equation}
which, on integrating, gives
\begin{equation} \label{D2t7}
 D^2(t) \leq \left({\mathrm{tr}}\rho^2\right)
 \left( e^{4\Gamma_0t} - 1 \right) + D^2(0)e^{4\Gamma_0t}.
\end{equation}
This inequality provides a strict upper bound of the fluctuations
of the random process. We note that the right-hand side of
(\ref{D2t7}) is finite for any finite time $t$. This leads to the
important conclusion that the fluctuations of the process are
finite for all finite times.

Let us discuss in more detail the case of a sharp initial state,
that is $D^2(0)=0$. We observe that for small times satisfying
$4\Gamma_0t \ll 1$ the root mean square distance then increases at
most as the square root of time,
\begin{equation}
 D(t) \leq \sqrt{({\mathrm{tr}}\rho^2)4\Gamma_0t}.
\end{equation}
For large times, $4\Gamma_0t \gg 1$, the root mean square distance
may increase, however, exponentially with time,
\begin{equation} \label{DT-INEQ}
 D(t) \leq \sqrt{{\mathrm{tr}}\rho^2} e^{2\Gamma_0t}.
\end{equation}
This shows that the stochastic method is useful for short and
intermediate times, where the relevant time scale is given by
$1/2\Gamma_0$. One further expects that the method is, in general,
not efficient numerically for times which are large compared to
$1/2\Gamma_0$, because of a possible exponential increase of the
fluctuations in this regime. It must be emphasized, however, that
the statistical errors can be reduced considerably by employing
the statistical independence of the increments
$|d\Phi_{\nu}\rangle$ (see Sec.~\ref{OBSERVABLES}), or by using a
more complicated ansatz for the structure of the stochastic states
(see Sec.~\ref{CONCLU}). It should also be noted that the
statistical errors are often much smaller than the upper bound
given in the inequality (\ref{DT-INEQ}). An example will be
discussed in Sec.~\ref{STOCHSIMSPIN}.

\subsection{The stochastic simulation method}

\subsubsection{Numerical algorithm} \label{ALGORITHM}
The stochastic simulation method consists in a numerical Monte
Carlo simulation of the stochastic differential equations
(\ref{STOCH1}) and (\ref{STOCH2}). A realizations $\psi_{\nu}(t)$,
$\chi_{\nu}(t)$ of the process can be generated by means of the
following algorithm.

1. Suppose that the last jump into states $\psi_{\nu}(t)$,
$\chi_{\nu}(t)$ occurred at some time $t$. In the case that $t$ is
the initial time $t=0$, these states are taken to be the
initial states which must be drawn from the probability
distribution representing the initial density matrix through
$\rho(0)={\mathrm{E}}(R(0))$.

2. The next jump takes place at time $t+\tau$, where the $\tau$ is
a stochastic time step, the random waiting time, which is to be
determined from the cumulative waiting time distribution function
\begin{equation} \label{WDF}
 F(\tau) = 1-\exp\left(-\int_{t}^{t+\tau}ds\Gamma_{\nu}(s)\right).
\end{equation}
A random number $\tau$ following this distribution can be
generated, for example, by drawing a uniform random number $\eta
\in (0,1)$ and by solving the equation
\begin{equation} \label{DEFTAU}
 \eta = \exp\left(-\int_{t}^{t+\tau}ds\Gamma_{\nu}(s)\right)
\end{equation}
for $\tau$. In between the previous and the next jump, that is
within the time interval $[t,t+\tau]$ the realization follows the
deterministic drift which is given by
\begin{eqnarray}
 \psi_{\nu}(t') &=& \psi_{\nu}(t), \label{DRIFTpsi}\\
 \chi_{\nu}(t') &=& \chi_{\nu}(t)
 \exp\left( \int_{t}^{t'}ds\Gamma_{\nu}(s) \right), \label{DRIFTchi}
\end{eqnarray}
where $t \leq t' \leq t+\tau$.

3. Select a particular jump, that is select a particular value of
the index $\alpha$ with probability
\begin{equation} \label{DEFALPHA}
 p_{\alpha\nu} =
 \frac{\Gamma_{\alpha\nu}(t+\tau)}{\sum_{\alpha}\Gamma_{\alpha\nu}(t+\tau)}.
\end{equation}
The corresponding jumps of the state vectors at time $t+\tau$ then
amount to the replacements
\begin{eqnarray}
 \psi_{\nu} &\longrightarrow&
 \frac{-i||\psi_{\nu}||}{||A_{\alpha}\psi_{\nu}||}
 A_{\alpha}\psi_{\nu}, \label{DEFJUMPS1} \\
 \chi_{\nu} &\longrightarrow&
 \frac{||\chi_{\nu}||}{||B_{\alpha}\chi_{\nu}||}
 B_{\alpha}\chi_{\nu}. \label{DEFJUMPS2}
\end{eqnarray}

Repeating these three steps until the desired final time $t_f$ is
reached on obtains a realization $\psi_{\nu}(t)$, $\chi_{\nu}(t)$
of the process over the whole time interval $[0,t_f]$. An
important feature of this algorithm is that it works with a random
time step the size of which is adapted automatically by the
algorithm: For large rates the time steps become small, while
small rates lead to an enhancement of the time steps. For example,
if $\Gamma_{\nu}$ is independent of time we simply have
\begin{equation} \label{LNETA}
 \tau = -\frac{1}{\Gamma_{\nu}} \ln \eta.
\end{equation}
In the case of a time-dependent rate $\Gamma_{\nu}(t)$ it may well
happen that the exponent in Eq.~(\ref{DEFTAU}) is bounded from
below and that, therefore, the exponential function converges to a
finite value $q>0$ as $\tau$ goes to infinity. For such a case one
distinguishes two cases. For $\eta > q$ one determines $\tau$ from
Eq.~(\ref{DEFTAU}), while for $\eta < q$ one sets $\tau = \infty$
in which case there will be no further jumps. An example of this
latter case will be shown in Sec.~\ref{JC-RES}.

Finally we remark that for a numerical implementation of the
simulation algorithm it might be more convenient to employ a PDP
with time-independent rates $\Gamma_{\nu}$. To this end one
replaces the stochastic differential equations (\ref{STOCH1}) and
(\ref{STOCH2}) by
\begin{eqnarray}
 d\psi_{\nu} &=& \sum_{\alpha}
 \left( \frac{-iA_{\alpha}}{\sqrt{\Gamma_{\alpha\nu}}}-I\right)
 \psi_{\nu} dN_{\alpha\nu}(t), \label{STOCH3} \\
 d\chi_{\nu} &=& \Gamma_{\nu} \chi_{\nu} dt
 + \sum_{\alpha}
 \left( \frac{B_{\alpha}}{\sqrt{\Gamma_{\alpha\nu}}}-I\right)
 \chi_{\nu} dN_{\alpha\nu}(t), \label{STOCH4}
\end{eqnarray}
with an appropriate choice for constant rates
$\Gamma_{\alpha\nu}$. The advantage of this method is that the
random waiting time is then always given by the simple expression
(\ref{LNETA}). The size of the statistical fluctuations, however,
can depend considerably on the choice of the $\Gamma_{\alpha\nu}$.

\subsubsection{Estimation of observables} \label{OBSERVABLES}
Suppose one has generated, by means of the algorithm described
above, a sample consisting of ${\mathcal{N}}$ realizations of the
process labeled by an index $r$,
\begin{equation}
 |\Phi_{\nu}^r(t)\rangle =
 \psi_{\nu}^r(t) \otimes \chi_{\nu}^r(t),
 \qquad r=1,2,\ldots,\mathcal{N}.
\end{equation}
The quantum expectation value
\begin{equation} \label{O-MEAN}
 {\mathcal{O}}(t) = {\mathrm{tr}} \{ \hat{\mathcal{O}}\rho(t) \}
 = {\mathrm{E}} \left( \langle\Phi_2(t)|\hat{\mathcal{O}}
 |\Phi_1(t)\rangle \right)
\end{equation}
of an observable $\hat{\mathcal{O}}$ of the total system can then
be estimated with the help of the ensemble average
\begin{equation} \label{O-MEAN-EST-1}
 {\mathcal{O}}_1(t) = \frac{1}{{\mathcal{N}}} \sum_r
 \langle\Phi_2^r(t)|\hat{\mathcal{O}}
 |\Phi_1^r(t)\rangle.
\end{equation}
In view of Eq.~(\ref{RHOS}) the reduced system's density matrix
$\rho_S(t)$ is given through the ensemble mean
\begin{equation} \label{OS-MEAN}
 \rho_S(t) = \frac{1}{{\mathcal{N}}} \sum_r
 |\psi_1^r(t)\rangle\langle\psi_2^r(t)|
 \langle \chi_2^r(t) | \chi_1^r(t) \rangle.
\end{equation}

As emphasized already, the $|\Phi_{\nu}(t)\rangle$ evolve
independently. Thus, if $|\Phi_1(0)\rangle$ and
$|\Phi_2(0)\rangle$ are independent, as it is the case for a sharp
initial value, for example, the processes $|\Phi_1(t)\rangle$ and
$|\Phi_2(t)\rangle$ are statistically independent. This implies
that Eq.~(\ref{O-MEAN}) can also be written in the following
equivalent way,
\begin{equation} \label{O-MEAN2}
 {\mathcal{O}}(t) = \langle\Psi_2(t)|\hat{\mathcal{O}}|\Psi_1(t)\rangle,
\end{equation}
where $|\Psi_{\nu}(t)\rangle={\mathrm{E}}(|\Phi_{\nu}(t)\rangle)$.
This suggests estimating the quantum expectation value
(\ref{O-MEAN}) by means of the alternative expression
\begin{equation} \label{O-MEAN-EST-2}
 {\mathcal{O}}_2(t) = \frac{1}{{\mathcal{N}}^2} \sum_{r,r'}
 \langle\Phi_2^r(t)|\hat{\mathcal{O}}
 |\Phi_1^{r'}(t)\rangle.
\end{equation}
Of course, the formulae (\ref{O-MEAN-EST-1}) and
(\ref{O-MEAN-EST-2}) lead to the same results in the limit of an
infinite number of realizations. However, for a finite sample the
statistical errors may differ considerably.

To illustrate the difference between the statistical estimates
given by (\ref{O-MEAN-EST-1}) and (\ref{O-MEAN-EST-2}), it
suffices to consider the case
$\hat{\mathcal{O}}=|\varphi\rangle\langle\varphi|$, where
$|\varphi\rangle$ may be any fixed state of the total system. We
introduce the random quantities $a=\langle \varphi | \Phi_1
\rangle$ and $b=\langle \varphi | \Phi_2 \rangle$, as well as the
corresponding realizations $a_r=\langle \varphi | \Phi_1^r
\rangle$ and $b_r=\langle \varphi | \Phi_2^r \rangle$. Equation
(\ref{O-MEAN-EST-1}) can then be written as
\begin{equation} \label{METHOD1}
 {\mathcal{O}}_1 = \frac{1}{{\mathcal{N}}} \sum_r b_r^{\ast}a_r.
\end{equation}
The corresponding statistical error is provided by the expression
\begin{equation} \label{DELTA1}
 \sigma_1 = \sqrt{\frac{{\mathrm{Var}}(a)}{\mathcal{N}}}
 \sqrt{{\mathrm{Var}}(a)+2|{\mathrm{E}}(a)|^2},
\end{equation}
where
\begin{equation} \label{VARA}
 {\mathrm{Var}}(a) \equiv {\mathrm{E}}(a^{\ast}a)
 - |{\mathrm{E}}(a)|^2
\end{equation}
is the variance of $a$, which is equal to the variance of $b$.

On the other hand, Eq.~(\ref{O-MEAN-EST-2}) leads to the
expression
\begin{equation} \label{METHOD2}
 {\mathcal{O}}_2 = \frac{1}{{\mathcal{N}}^2}
 \sum_{r,r'} b_r^{\ast}a_{r'}.
\end{equation}
The usage of this formula for the estimation of ${\mathcal{O}}$ is
more efficient, in general, since the corresponding statistical
error
\begin{equation} \label{DELTA2}
 \sigma_2 = \sqrt{\frac{{\mathrm{Var}}(a)}{\mathcal{N}}}
 \sqrt{2|{\mathrm{E}}(a)|^2}
\end{equation}
is smaller than $\sigma_1$. The second method based on
Eq.~(\ref{METHOD2}) is thus to be preferred since it yields
considerably smaller fluctuations. This difference between both
methods becomes particularly important if $|{\mathrm{E}}(a)|^2$,
the quantity to be estimated, is small. The simulations presented
in Sec.~\ref{JC-RES} and Sec.~\ref{DET-RES}, for example, have
been carried out using this second method.

\subsubsection{Quantum correlation functions} \label{CORRFUNC}
The fact that the stochastic method involves a pair of random
wave functions also enables the design of an exact method for the
determination of multitime correlation functions. The underlying
idea is similar to the one employed in \cite{MULTITIME} for the
calculation of correlation functions of quantum Markov processes.

We restrict the discussion to the case of an arbitrary two-time
correlation function of the form $\langle X(t) Y(0) \rangle$. In
the interaction picture we can write (assuming $t\geq 0$)
\begin{eqnarray} \label{DEF-CORR}
 \langle X(t) Y(0) \rangle &=& {\mathrm{tr}} \left(
 X(t) U(t) Y(0) \rho(0) U^{\dagger}(t) \right) \nonumber \\
 &=& {\mathrm{E}} \left(
 \langle \Phi_2(t) | X(t) | \Phi^Y_1(t) \rangle \right),
\end{eqnarray}
where $X(t)$ and $Y(t)$ are arbitrary operators in the interaction
picture, and $U(t)$ denotes the interaction picture time-evolution
operator of the total system over time $t$. The second line in
Eq.~(\ref{DEF-CORR}) provides the stochastic representation of the
quantum correlation function. In this expression both
$|\Phi^Y_1(t)\rangle$ and $|\Phi_2(t)\rangle$ follow the
stochastic dynamics developed in Sec.~\ref{PDP}. However, while
the initial state of $|\Phi_2(t)\rangle$ is $|\Phi_2(0)\rangle$,
the stochastic process $|\Phi^Y_1(t)\rangle$ evolves from the new
initial state $|\Phi^Y_1(0)\rangle=Y(0)| \Phi_1(0)\rangle$. With
this modification the stochastic algorithm for the determination
of the correlation function is the same as above. The method can
easily be generalized to the case of multitime correlation
functions. An example will be studied in Sec.~\ref{JC-RES}.

\section{Decay into a bosonic reservoir}
\label{SECBOSONIC}

To illustrate the general method developed in Sec.~\ref{GENERAL}
we first study the model of a two-state system with excited state
$|e\rangle$, ground state $|g\rangle$, and corresponding
transition frequency $\omega_0$. This system is coupled to a
bosonic reservoir consisting of field modes which will be labeled
by an index $k$. The corresponding field operators that annihilate
and create particles of frequency $\omega_k$ are denoted by $b_k$
and $b_k^{\dagger}$, respectively. The interaction picture
Hamiltonian is taken to be of the form
\begin{equation} \label{H-I-BOSONIC}
 H_I(t) = \sigma_+ B(t) + \sigma_- B^{\dagger}(t).
\end{equation}
The operators $\sigma_+=|e\rangle\langle g|$ and
$\sigma_-=|g\rangle\langle e|$ are the raising and lowering
operators of the two-state system, while the reservoir operator
$B(t)$ is given by
\begin{equation}
 B(t) = \sum_k g_k b_k e^{i(\omega_0-\omega_k)t},
\end{equation}
with mode-dependent coupling constants $g_k$. As a simple example
we investigate the initial state
\begin{equation} \label{INIT-BOSONIC}
 |\Phi_{\nu}(0)\rangle = \psi(0) \otimes \chi(0)
 = |e\rangle \otimes |0\rangle,
\end{equation}
where $|0\rangle$ denotes the vacuum state of the reservoir. This
initial state is statistically sharp and corresponds to the
density matrix $\rho(0)=|e\rangle\langle e|\otimes|0\rangle\langle
0|$ of the total system. This model can be solved analytically.
The central physical quantity that determines the influence of the
reservoir modes on the reduced system dynamics is provided by the
bath correlation function
\begin{eqnarray}
 f(t'-t) &=& \langle 0 | B(t') B^{\dagger}(t) | 0 \rangle \\
 &=& \int d\omega J(\omega) \exp[i(\omega_0-\omega)(t'-t)],
 \nonumber
\end{eqnarray}
which has been expressed here in terms of the spectral density
$J(\omega)$.

\subsection{Description of the algorithm} \label{SIM-BOSONIC}
In the notation of Sec.~\ref{PDP} we have $\alpha=1,2$ and
$A_1=\sigma_+$, $A_2=\sigma_-$, $B_1(t)=B(t)$ and
$B_2(t)=B^{\dagger}(t)$. The application of the general technique
of Sec.~\ref{ALGORITHM} to the present case leads to the following
algorithm of simulating the stochastic dynamics.

{\em{After an even number of jumps}} the reservoir state
$\chi_{\nu}$ is proportional to the vacuum state. We thus infer
from Eq.~(\ref{DEFGANU}) that the transition rates are given by
\begin{equation}
 \Gamma_{\nu}(t')
 = \frac{||B^{\dagger}(t')\chi_{\nu}(t')||}{||\chi_{\nu}(t')||}
 = ||B^{\dagger}(t')|0\rangle|| = \sqrt{f(0)}.
\end{equation}
Since these rates are constant in time the random time step $\tau$
is determined by Eq.~(\ref{LNETA}), that is
$\tau=-\ln\eta/\sqrt{f(0)}$ with a uniform random number $\eta$ in
the interval $(0,1)$. Suppose that the previous jump took place at
time $t$. Over the time interval $[t,t+\tau]$ the state
$\chi_{\nu}$ then changes continuously according to
\begin{equation}
 \chi_{\nu}(t') = \chi_{\nu}(t) e^{\Gamma_{\nu}\cdot (t'-t)},
 \qquad t \leq t' \leq t+\tau,
\end{equation}
until at time $t+\tau$ the jumps described in
Eqs.~(\ref{DEFJUMPS1}) and (\ref{DEFJUMPS2}) occur,
\begin{eqnarray}
 \psi_{\nu}(t+\tau) &\longrightarrow&
 -i\sigma_-\psi_{\nu}(t+\tau), \\
 \chi_{\nu}(t+\tau) &\longrightarrow&
  \frac{B^{\dagger}(t+\tau)}{\sqrt{f(0)}} \chi_{\nu}(t+\tau).
\end{eqnarray}
Note, in particular, that $\chi_{\nu}$ jumps into a 1-particle
state.

{\em{After an odd number of jumps}} the reservoir state
$\chi_{\nu}$ represents a 1-particle state which was created out
of the field vacuum at the time $t$ of the last jump. Invoking
again Eq.~(\ref{DEFGANU}) we find that the transition rates are
now given by
\begin{eqnarray} \label{RATES-JC}
 \Gamma_{\nu}(t')
 &=& \frac{||B(t')\chi_{\nu}(t')||}{||\chi_{\nu}(t')||}
 =\frac{||B(t')B^{\dagger}(t)|0\rangle||}{||B^{\dagger}(t)|0\rangle||}
 \nonumber \\
 &=& \frac{|f(t'-t)|}{\sqrt{f(0)}}.
\end{eqnarray}
We observe that these rates are time-dependent such that the
random time step $\tau$ as well as the deterministic drift of
$\chi_{\nu}$ must be determined from Eq.~(\ref{DEFTAU}) and
(\ref{DRIFTchi}), respectively. In the present case we thus have
\begin{equation} \label{WDF-JC}
 \eta = \exp\left(-\int_{0}^{\tau}ds |f(s)|/\sqrt{f(0)} \right),
\end{equation}
and
\begin{equation}
 \chi_{\nu}(t') = \chi_{\nu}(t)
 \exp\left( \int_{0}^{t'-t}ds |f(s)|/\sqrt{f(0)} \right).
\end{equation}
Finally, the jumps at time $t+\tau$ take the form:
\begin{eqnarray}
 \psi_{\nu}(t+\tau) &\longrightarrow&
 -i\sigma_+\psi_{\nu}(t+\tau), \label{TRANS1} \\
 \chi_{\nu}(t+\tau) &\longrightarrow&
  \frac{\sqrt{f(0)}}{|f(\tau)|} B(t+\tau) \chi_{\nu}(t+\tau).
 \label{TRANS2}
\end{eqnarray}
At time $t+\tau$ the environment thus jumps back into a state
which is proportional to the vacuum state. In terms of
$\tilde{\chi}_{\nu}(t)$, which is defined to be the reservoir
state just {\textit{before}} the previous jump at time $t$, we can
write the transition (\ref{TRANS2}) as
\begin{equation} \label{TRANS3}
 \chi_{\nu}(t+\tau) \longrightarrow
 \frac{f(\tau)}{|f(\tau)|} \tilde{\chi}_{\nu}(t)
 \exp \left( \int_{0}^{\tau}ds |f(s)|/\sqrt{f(0)} \right).
\end{equation}
This algorithm will be applied in the following two sections to
the damped Jaynes-Cummings model on resonance and with a finite
detuning.

\subsection{Damped Jaynes-Cummings model on resonance} \label{JC-RES}
The spectral density of the damped Jaynes-Cummings model on
resonance is given by
\begin{equation} \label{SPECDENS1}
 J(\omega)=\frac{1}{2\pi}\frac{\gamma_0\lambda^2}
 {(\omega_{0}-\omega)^2+\lambda^2},
\end{equation}
which yields the bath correlation function
\begin{equation} \label{CORRFUNC1}
 f(t'-t) = \frac{1}{2} \gamma_0 \lambda e^{-\lambda |t'-t|}.
\end{equation}
This model can be used to describe the coupling of a two-level
atom to an electromagnetic cavity mode which in turn is coupled to
the continuum of modes of the electromagnetic field vacuum. The
quantity $\lambda^{-1}$ is the correlation time of the reservoir,
while $\gamma_0^{-1}$ can be interpreted as the Markovian
relaxation time of the open system.

\begin{figure}[htb]
\includegraphics[width=\linewidth]{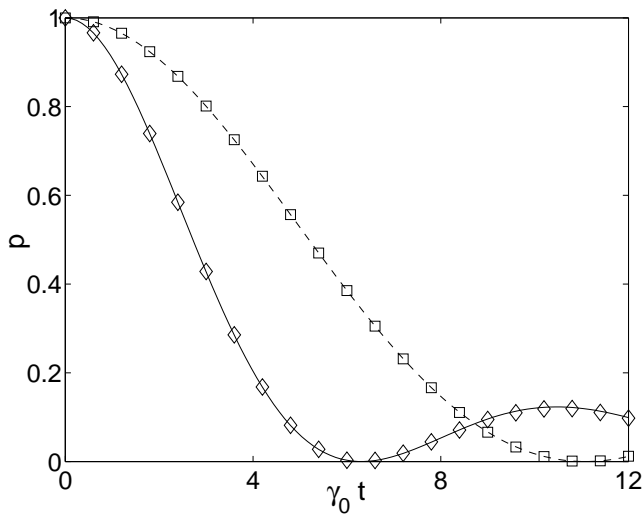}
\caption{\label{figure1}
 Excited state probability $p(t)$
 (Eq.~(\ref{def-p-t})) of the damped Jaynes-Cummings model.
 Symbols: Monte Carlo simulations of the stochastic differential
 equations (\ref{STOCH1}) and (\ref{STOCH2}) with
 ${\mathcal{N}} = 5 \cdot 10^6$ realizations for the parameters
 $\lambda^{-1} = 5\gamma_0^{-1}$ (diamonds) and $\lambda^{-1} =
 20\gamma_0^{-1}$ (squares). The corresponding analytical solutions
 are given by the continuous and the broken line.}
\end{figure}

The application of the simulation algorithm detailed in
Sec.~\ref{SIM-BOSONIC} to this situation is straightforward. In
particular, we note that according to Eqs.~(\ref{RATES-JC}) and
(\ref{CORRFUNC1}) the waiting time distribution (\ref{WDF}) after
an odd number of jumps takes the form
\begin{equation}
 F(\tau) = 1 - \exp \left( -\sqrt{\frac{\gamma_0}{2\lambda}}
 \left[1-e^{-\lambda \tau} \right] \right).
\end{equation}
Hence, the probability that no further jumps occur equals
\begin{equation}
 q = 1-\lim_{\tau \rightarrow \infty} F(\tau) =
 \exp \left( -\sqrt{\frac{\gamma_0}{2\lambda}} \right).
\end{equation}
This means that in the case $\eta < q$ no further jumps occur,
while in the case $\eta > q$ the random time step is determined by
Eq.~(\ref{WDF-JC}) which yields
\begin{equation}
 \tau = -\frac{1}{\lambda} \ln \left( 1 +
 \sqrt{\frac{2\lambda}{\gamma_0}} \ln \eta \right).
\end{equation}

\begin{figure}[htb]
\includegraphics[width=\linewidth]{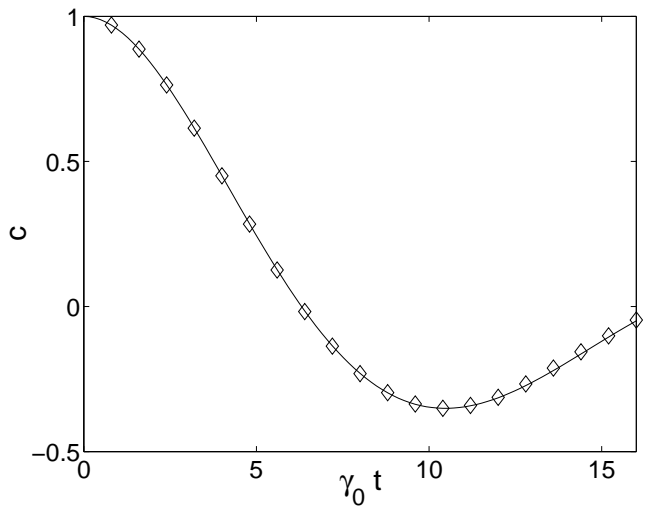}
\caption{\label{figure2}
 The correlation function $c(t)$ (Eq.~(\ref{def-c-t}))
 of the damped Jaynes-Cummings model: Analytical solution
 (continuous line) and Monte Carlo simulation of the stochastic
 differential equations (\ref{STOCH1}) and (\ref{STOCH2}) (diamonds) for
 $\lambda^{-1} = 5\gamma_0^{-1}$ and ${\mathcal{N}}=10^7$ realizations.}
\end{figure}

Results of Monte Carlo simulations of the damped Jaynes-Cummings
model are presented in Fig.~\ref{figure1}, which shows the
population of the excited state,
\begin{equation} \label{def-p-t}
 p(t) = {\mathrm{E}}\left( \langle e |\psi_1\rangle\langle
 \psi_2| e\rangle \langle \chi_2 | \chi_1 \rangle \right),
\end{equation}
estimated from a sample of realizations of the stochastic process
using the estimator described by Eq.~(\ref{METHOD2}). As can be
seen from the figure, the simulation results reproduce the
analytical curves with high accuracy. We note that for the
parameter values chosen the reservoir correlation time
$\lambda^{-1}$ is larger than the reduced system's Markovian
relaxation time $\gamma_0^{-1}$. We therefore observe a pronounced
non-Markovian behavior and large deviations form the Born-Markov
dynamics. For small and intermediate couplings, the open system
dynamics derived from the model described by the interaction
Hamiltonian (\ref{H-I-BOSONIC}) and initial conditions
(\ref{INIT-BOSONIC}) satisfies a time-local master equation of the
form $\dot{\rho}_S(t)={\mathcal{K}}(t)\rho_S(t)$ with a
time-dependent super-operator ${\mathcal{K}}(t)$. However, the TCL
expansion of the generator ${\mathcal{K}}(t)$ breaks down in the
strong coupling regime given by $\lambda^{-1} >
\frac{1}{2}\gamma_0^{-1}$ for times $t > t_0$, where $t_0$ denotes
the first positive zero of $p(t)$. Beyond the singularity at
$t=t_0$ the TCL expansion of the master equation is therefore not
capable of describing the reduced system dynamics which develops a
long memory time of the order $t_0$. However, as is exemplified in
the figure, the stochastic simulation is seen to describe
correctly the full non-Markovian behavior of the reduced system
even in the strong coupling regime.

To give an example of the simulation of correlation functions we
investigate the quantity $\langle \sigma_+(t) \sigma_-(0) \rangle$
which can be determined with the help of the method described in
Sec.~\ref{CORRFUNC}. Figure \ref{figure2} shows the simulation
results for the quantity
\begin{equation} \label{def-c-t}
 c(t) \equiv e^{-i\omega_0t} \langle \sigma_+(t) \sigma_-(0) \rangle,
\end{equation}
which again nicely fit the analytical curve.

\subsection{Jaynes-Cummings model with detuning} \label{DET-RES}
If the cavity mode is detuned from the atomic transition frequency
by an amount $\Delta$ the spectral density becomes
\begin{equation}
 J(\omega)=\frac{1}{2\pi}\frac{\gamma_0\lambda^2}
 {(\omega_{0}-\Delta-\omega)^2+\lambda^2},
\end{equation}
which leads to the reservoir correlation function
\begin{equation} \label{CORRFUNC2}
 f(t'-t) = \frac{1}{2} \gamma_0 \lambda e^{i\Delta (t'-t)
 - \lambda |t'-t|}.
\end{equation}
We can again use the simulation algorithm described in
Sec.~\ref{SIM-BOSONIC}, although, by contrast to the previous
case, the correlation function (\ref{CORRFUNC2}) is
complex-valued. Since the transition rates and the deterministic
drift of the process depend on the absolute value of $f$, the only
modification of the algorithm for the resonant case appears in
Eq.~(\ref{TRANS3}) which describes the even jumps into the vacuum
state.

\begin{figure}[htb]
\includegraphics[width=\linewidth]{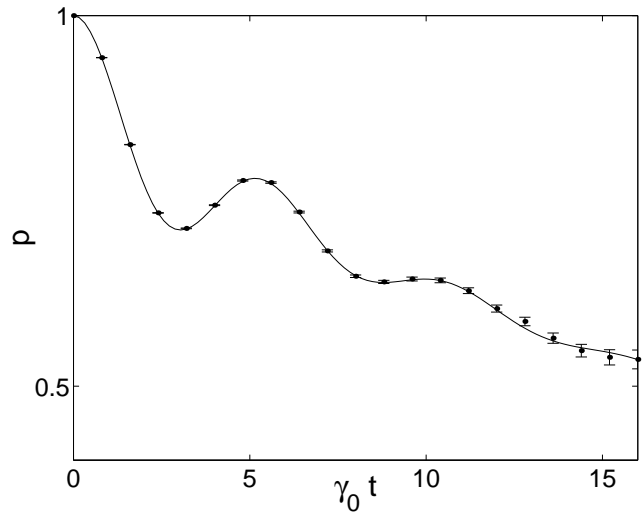}
\caption{\label{figure3}
 The excited state probability $p(t)$ (Eq.~(\ref{def-p-t}))
 of the damped Jaynes-Cummings model with detuning:
 Analytical solution (continuous line) and Monte Carlo simulation
 of the stochastic differential equations (\ref{STOCH1}) and
 (\ref{STOCH2}) (dots and errorbars) for
 $\lambda^{-1} = 5\gamma_0^{-1}$, $\Delta = \gamma_0$ and
 ${\mathcal{N}}=10^7$ realizations.}
\end{figure}

An example of the simulation results is shown in
Fig.~\ref{figure3}. The detuning $\Delta$ influences both the
coherent dynamics of the system as well as the dissipation
mechanism. This leads to a slower decay and to an oscillatory
behavior of the excited state probability, which is correctly
reproduced by the stochastic simulation.

\section{Interaction with a spin bath} \label{SPINBAD}
The stochastic method developed in Sec.~\ref{GENERAL} is not
restricted to the treatment of bosonic reservoirs. It is also
applicable to the dynamics of open systems coupled to spin
environments. As an example, we examine here a specific central
spin model which may be used to model the interaction of a single
electron spin confined to a quantum dot with a bath of nuclear
spins \cite{LOSS}.

\subsection{Description of the model}
The model is defined by the total Hamiltonian
\begin{equation}
 H = \frac{\omega_0}{2} \sigma_3 +
 \sum_{j=1}^{N} A^{(j)} \vec{\sigma}\cdot\vec{\sigma}^{(j)}.
\end{equation}
The central spin is represented by the Pauli spin operator
$\vec{\sigma}$, while the $N$ bath spins are given by the spin
operators $\vec{\sigma}^{(j)}$ with $j=1,2,\ldots,N$. The coupling 
of the central spin to the $j$th bath spin is described by the 
constant $A^{(j)}$. For simplicity, the coupling constants are taken to be
$A^{(j)}=A/\sqrt{N}$. The corresponding interaction picture
Hamiltonian can be written as
\begin{equation} \label{HINT-SPIN}
 H_I(t) = \sigma_3 B_3(t) + \sigma_+ B_-(t) + \sigma_- B_+(t)
\end{equation}
with
\begin{eqnarray}
 B_3 &=& \sum_j A^{(j)} \sigma_3^{(j)}, \\
 B_{\pm} &=& \sum_j 2A^{(j)} \sigma_{\pm}^{(j)} e^{\mp i\omega_0t}.
 \label{BPM}
\end{eqnarray}

Our aim is to determine the coherence of the central spin,
\begin{equation} \label{COH-SPIN-1}
 \rho_{+-}(t) = \langle +|\rho_S(t)|-\rangle,
\end{equation}
where $|\pm\rangle$ are the eigenstates of the 3-component
$\sigma_3$ of the central spin $\vec{\sigma}$ with eigenvalues
$\pm 1$. Within the stochastic simulation technique this quantity
is represented through the expectation value (see
Eq.~(\ref{RHOS}))
\begin{equation} \label{COH-SPIN-2}
 \rho_{+-}(t) = {\mathrm{E}}
 \left( \langle + | \psi_+ \rangle \langle \psi_- | - \rangle
 \langle \chi_- | \chi_+ \rangle \right),
\end{equation}
where we write here
$|\Phi_{\nu}\rangle\equiv|\Phi_{\pm}\rangle=\psi_{\pm}\otimes\chi_{\pm}$
for the stochastic states, that is the index $\nu$ takes on the
values $\nu=\pm$. The initial state is taken to be
\begin{equation} \label{INIT-SPIN}
 \rho(0) = |+\rangle\langle-| \otimes \frac{1}{2^N}I_E.
\end{equation}
$I_E$ denotes the unit matrix in the $2^N$-dimensional state space
${\mathcal{H}}_E$ of the spin bath. The spin bath is thus in an
unpolarized initial state.

\subsection{Simulation algorithm and results} \label{STOCHSIMSPIN}
To apply the simulation technique it is useful to realize the
unpolarized initial state $2^{-N}I_E$ of the spin bath with the
help of an appropriate set of basis states of the Hilbert space
${\cal{H}}_E$ spanned by the $N$ bath spins. To this end, we
introduce states $|j,m\rangle$ which are defined as simultaneous
eigenstates of the square $\vec{J}\,^2$ of the total spin angular
momentum $\vec{J}$ of the bath and of its 3-component $J_3$. The
initial state can then be represented by
\begin{equation} \label{INIT-SPIN-JM}
 |\Phi_{\pm}(0)\rangle = |\pm\rangle \otimes |j,m\rangle
\end{equation}
with an appropriate probability distribution of the corresponding
quantum numbers $j$ and $m$ which will be constructed below.

The state $|\Phi_{\pm}(0)\rangle$ defined in (\ref{INIT-SPIN-JM})
is an eigenstate of the 3-component $\frac{1}{2}\sigma_3+J_3$ of
the total spin angular momentum, which is a conserved quantity,
corresponding to the eigenvalue $\frac{1}{2}(\pm 1 +2m)$. This
fact enables us to carry out the canonical transformation
$|\Phi_{\pm}(t)\rangle\longrightarrow|\tilde{\Phi}_{\pm}(t)\rangle$
defined by
\begin{equation} \label{TRAFO-SPIN}
 |\Phi_{\pm}(t)\rangle = \exp\left[ \frac{-iAt}{\sqrt{N}}
 \left( (\pm 1+2m)\sigma_3 -1 \right) \right]
 |\tilde{\Phi}_{\pm}(t)\rangle,
\end{equation}
which transforms the interaction Hamiltonian (\ref{HINT-SPIN})
into
\begin{equation} \label{TILDEHINT-SPIN}
 \tilde{H}_I(t) = \sigma_+ B_-(t) + \sigma_- B_+(t).
\end{equation}
In this equation the $B_{\pm}(t)$ are given again by
Eq.~(\ref{BPM}), where, however, $\omega_0$ must be replaced by
the new frequencies $\omega_{\pm}$:
\begin{equation} \label{OMEPM}
 \omega_0 \longrightarrow
 \omega_{\pm} = \omega_0 + \frac{2A}{\sqrt{N}}(\pm 1 + 2m).
\end{equation}
In terms of the stochastic states
$|\tilde{\Phi}_{\pm}\rangle=\tilde{\psi}_{\pm}\otimes\tilde{\chi}_{\pm}$
the coherence of the central spin is then given by the expectation
value
\begin{equation} \label{COHERENCE-SPIN}
 \rho_{+-}(t) = {\mathrm{E}}
 \left( e^{-4iAmt/\sqrt{N}}
 \langle + | \tilde{\psi}_+ \rangle \langle \tilde{\psi}_- | - \rangle
 \langle \tilde{\chi}_- | \tilde{\chi}_+ \rangle \right).
\end{equation}
Summarizing, we can simulate, employing the method developed in
Sec.~\ref{GENERAL}, the stochastic dynamics corresponding to the
new interaction Hamiltonian (\ref{TILDEHINT-SPIN}) and estimate
the coherence by means of the formula (\ref{COHERENCE-SPIN}). The
canonical transformation (\ref{TRAFO-SPIN}) is accounted for in
this formula by the exponential factor $\exp[-4iAmt/\sqrt{N}]$.

In order to see more explicitly how the method works it may be
instructive at this point to consider first the simpler model
obtained by omitting the terms $\sigma_{\pm}B_{\mp}(t)$ of the
interaction Hamiltonian (\ref{HINT-SPIN}). The transformed
Hamiltonian (\ref{TILDEHINT-SPIN}) is then identically zero and
the expression (\ref{COHERENCE-SPIN}) for the coherence of the
central spin becomes
\begin{equation} \label{COHERENCE-SPIN-2}
 \rho_{+-}(t) = {\mathrm{E}} \left( e^{-4iAmt/\sqrt{N}} \right)
 = \sum_{m=-N/2}^{+N/2} p_m e^{-4iAmt/\sqrt{N}},
\end{equation}
where $p_m$ is the probability of finding a basis state with
quantum number $m$ in the unpolarized initial mixture. Since all
basis states are equally likely in this initial mixture, $p_m$ is
found to be
\begin{equation} \label{BINOM}
 p_m = \frac{1}{2^N} \binom{N}{\frac{N}{2} + m}.
\end{equation}
Here, $2^N$ is the total number of basis states of the bath of $N$
spins (the dimension of ${\mathcal{H}}_E$), while the binomial
coefficient counts the number of basis states corresponding to a
given value of $m$. The summation in Eq.~(\ref{COHERENCE-SPIN-2})
can easily be carried out to give
\begin{equation} \label{COHERENCE-SPIN-3}
 \rho_{+-}(t) = \left[ \cos\left(
 \frac{2At}{\sqrt{N}}\right)\right]^N,
\end{equation}
which is the exact expression for the coherence of the central
spin. We note that this expression may be approximated by
\begin{equation} \label{COHERENCE-SPIN-4}
 \rho_{+-}(t) = e^{-2A^2t^2}
\end{equation}
in the limit of a large number of bath spins, $N\longrightarrow
\infty$, showing an exponential decay of the coherence of the
central spin. Thus we see that the stochastic simulation for this
simplified model reduces to the generation of a binomially
distributed random number $m$ and to the estimation of the
expectation value (\ref{COHERENCE-SPIN-2}).

We turn again to the discussion of the full model described by the
Hamiltonian (\ref{HINT-SPIN}). Employing the method described
above and using the transformed interaction Hamiltonian
(\ref{TILDEHINT-SPIN}) we see that the simulation algorithm is
quite similar to the one used already in the bosonic case. In
fact, the simulation technique turns out to be even simpler.
Suppose we have drawn the initial state $|\pm\rangle \otimes
|j,m\rangle$. The bath state $\tilde{\chi}_{\pm}(t)$ then jumps
between states which are proportional to $|j,m\rangle$ and
$|j,m\pm 1\rangle$. The corresponding jump rate
\begin{equation} \label{RATES-SPIN}
 \Gamma_{\pm} = 2A\sqrt{\frac{j(j+1)-m(m\pm 1)}{N}}
\end{equation}
is independent of time. The waiting time of the PDP is therefore
always exponentially distributed, which makes the numerical
implementation particularly easy for this case. A detailed
analysis of the process reveals that the coherence can be
represented through the expectation value
\begin{eqnarray} \label{EXPEC-SPIN-COH}
 \rho_{+-}(t) &=& {\mathrm{E}} \big( e^{-4iAmt/\sqrt{N}}
 (-1)^{(k_++k_-)/2} e^{(\Gamma_++\Gamma_-)t} \nonumber \\
 &~& \;\;\;
 \times \exp(i\omega_+(\tau_2^++\tau_4^+\ldots+\tau_{k_+}^+))
 \nonumber \\
 &~& \;\;\; \times
 \exp(i\omega_-(\tau_2^-+\tau_4^-\ldots+\tau_{k_-}^-))
 \big).
\end{eqnarray}
Here, $\tau_{2n}^{\pm}$ denotes the random time step before the
$2n$th jump of $|\tilde{\Phi}_{\pm}\rangle$, while $\Gamma_{\pm}$
and $\omega_{\pm}$ have already been defined in
Eqs.~(\ref{RATES-SPIN}) and (\ref{OMEPM}). The quantity $k_{\pm}$
is defined as the total number of jumps of
$|\tilde{\Phi}_{\pm}(t)\rangle$ during the time interval from $0$
to $t$. The integers $k_{\pm}$ may be supposed to be even since
only trajectories with an even number of jumps contribute to the
expectation value (\ref{EXPEC-SPIN-COH}).

It remains to explain how to generate, in the general case, the
initial states $|j,m\rangle$ in Eq.~(\ref{INIT-SPIN-JM}). More
precisely, these states should be written as
$|\lambda,j,m\rangle$, where $\lambda$ stands for an additional
quantum number which, together with $j$ and $m$, uniquely fixes
the basis state. The quantum number $\lambda$ corresponds to
further observables of the spin bath which commute with
$\vec{J}\,^2$ and $J_3$. If $N$ is even $j$ takes on the values
$j=0,1,2,\ldots,\frac{N}{2}$, while
$j=\frac{1}{2},\frac{3}{2},\ldots,\frac{N}{2}$ if $N$ is odd. For
a given value of $j$ the quantum number $m$ takes on the values
$m=-j,-j+1,\ldots,+j$.

In order to achieve that the initial ensemble represents the
unpolarized bath state, that is
\begin{equation}
 {\mathrm{E}}(|\lambda,j,m\rangle\langle \lambda,j,m|)
 = \frac{1}{2^N} I_E,
\end{equation}
all basis states $|\lambda,j,m\rangle$ must occur with the same
probability of $2^{-N}$. Since the value of the quantum number
$\lambda$ is irrelevant in the simulation scheme, we need the
probability $P(j,m)$ of finding the pair of quantum numbers
$(j,m)$ in the initial ensemble. This probability can be written
as
\begin{equation} \label{P-j-m}
 P(j,m) = 2^{-N} a_j^N.
\end{equation}
The quantity $a_j^N$ denotes the number of times a given angular
momentum $j$ appears in the decomposition of the Hilbert space
${\mathcal{H}}_E$ of $N$ spins into irreducible subspaces of the
rotation group. Since a certain $j$-manifold consists of $(2j+1)$
states, distinguished by their values of the quantum number $m$,
we can also say that $(2j+1)a_j^N$ is equal to the number of
independent ways the $N$ bath spins can be coupled to give the
total angular momentum $j$. For example, the Hilbert space of
$N=4$ spins decomposes into two ($j=0$)-manifolds, three
($j=1$)-manifolds, and one ($j=2$)-manifold, that is we have
$a_0^4=2$, $a_1^4=3$, and $a_2^4=1$. It may be shown \cite{BOSE}
that $a_j^N$ is given by the general expression
\begin{equation} \label{A-J-N}
 a_j^N = \binom{N}{\frac{N}{2}+j}-\binom{N}{\frac{N}{2}+j+1}.
\end{equation}
We note that $P(j,m)$ is normalized,
\begin{equation}
 \sum_j \sum_{m=-j}^{+j} P(j,m) = 1,
\end{equation}
and does of course not depend on $m$. In summary, the quantum
numbers $(j,m)$ of the initial ensemble follow the distribution
$P(j,m)$ given by the expressions (\ref{P-j-m}) and (\ref{A-J-N}).
In the stochastic simulation algorithm one therefore has to
generate a sample of random numbers $(j,m)$ with this
distribution, which is easily done making use of the inversion
method, for example.

\begin{figure}[htb]
\includegraphics[width=\linewidth]{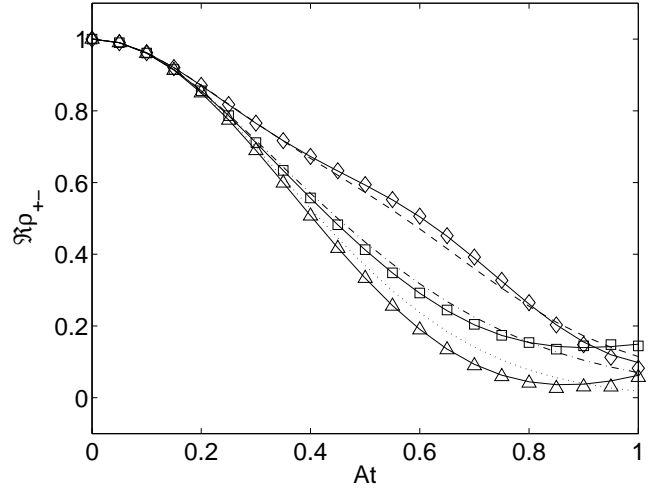}
\caption{\label{figure4}
 Real part of the coherence (\ref{COH-SPIN-1}) of the
 central spin interacting with a spin bath through the Hamiltonian
 (\ref{HINT-SPIN}) with $N=10^3$. Symbols: Monte Carlo simulation
 of the stochastic differential equations (\ref{STOCH1}) and
 (\ref{STOCH2}) using ${\mathcal{N}}=2\cdot 10^7$ realizations for
 the parameters $A/\omega_0=0.1$ (diamonds), $A/\omega_0=0.2$
 (squares), and $A/\omega_0=10$ (triangles). Continuous lines:
 Corresponding solutions of the von Neumann equation
 (\ref{NEUMANN}). The dashed line ($A/\omega_0=0.1$), the
 dashed-dotted line ($A/\omega_0=0.2$), and the dotted line
 ($A/\omega_0=10$) show the results obtained from the TCL master
 equation in second order (Eqs.~(\ref{TCL2}) and (\ref{GAMMA2})).}
\end{figure}

Examples of Monte Carlo simulations of the central spin model are
shown in Fig.~\ref{figure4}. One observes that the PDP reproduces
the von Neumann dynamics with high accuracy. We do not show
errorbars in the figure because the statistical errors are smaller
than the size of the symbols. The figure also displays the results
found with the help of the second-order TCL master equation of the
central spin which is given by
\begin{eqnarray}
 \frac{d}{dt}\rho_S &=& -2iA^2\frac{1-\cos\omega_0t}{\omega_0}
 [\sigma_3,\rho_S] \\
 &~& - A^2t[\sigma_3,[\sigma_3,\rho_S]] \nonumber \\
 &~& + 4A^2\frac{\sin\omega_0t}{\omega_0}
 \left( \sigma_-\rho_S\sigma_+
 - \frac{1}{2} \{ \sigma_+\sigma_-,\rho_S\} \right. \nonumber \\
 &~& \qquad \qquad \qquad \left. + \sigma_+\rho_S\sigma_-
 - \frac{1}{2} \{ \sigma_-\sigma_+,\rho_S\}
 \right). \nonumber
\end{eqnarray}
The solution of this master equation is easily constructed. It
yields the expression
\begin{equation} \label{TCL2}
 \rho_{+-}(t)=\exp[-\Gamma(t)]\rho_{+-}(0)
\end{equation}
for the coherence of the central spin, where
\begin{eqnarray} \label{GAMMA2}
 \Gamma(t) &=& \frac{4iA^2t}{\omega_0}
 \left( 1 - \frac{\sin\omega_0t}{\omega_0t} \right) \nonumber \\
 &~& + 2A^2t^2 \left( 1 + 2\frac{1-\cos\omega_0t}{(\omega_0t)^2}
 \right).
\end{eqnarray}
For the parameter values chosen the exact dynamics of the central
spin is seen to deviate significantly from the one predicted by
the second-order TCL master equation.

\begin{figure}[htb]
\includegraphics[width=\linewidth]{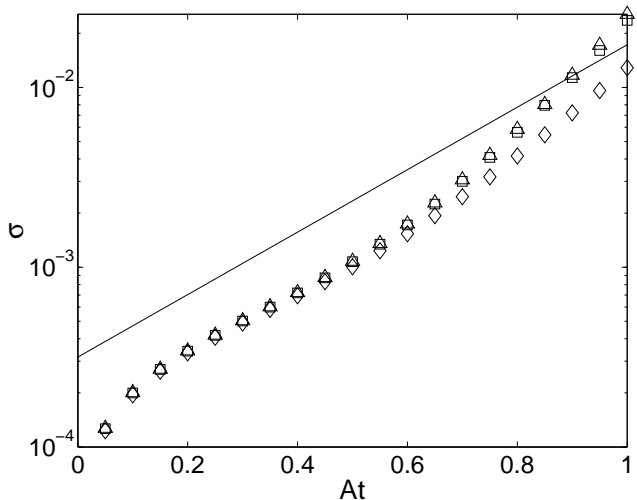}
\caption{\label{figure5}
 Statistical errors $\sigma(t)$ of Monte Carlo simulations
 of the central spin model with $10^7$ realizations, $A/\omega_0=0.5$
 and three different values of the number of bath spins: $N=10$
 (diamonds), $N=100$ (squares), and $N=1000$ (triangles). The
 continuous line shows the estimate given by Eq.~(\ref{ESTIMATE}).}
\end{figure}

Figure~\ref{figure5} presents an example of the behavior of the
fluctuations of the stochastic process. The figure shows a plot of
the statistical errors $\sigma(t)$ of three Monte Carlo
simulations with a fixed number ${\mathcal{N}}$ of realizations,
but with three different values of the number $N$ of bath spins.
We conclude from the figure that, within the range of time
investigated, $\sigma(t)$ is roughly independent of $N$. To
understand this behavior we refer to expression
(\ref{EXPEC-SPIN-COH}) which yields
\begin{equation}
 \sigma(t) \leq \sqrt{
 \frac{{\mathrm{E}}\left( \exp[2(\Gamma_++\Gamma_-)]
 \right)}{{\mathcal{N}}}}.
\end{equation}
The right-hand side of this inequality may be estimated by
replacing the random quantities $\Gamma_{\pm}$ by suitable
averages using the distribution (\ref{P-j-m}). This gives the
estimate
\begin{equation} \label{ESTIMATE}
 \sigma(t) \sim \frac{\exp[4At]}{\sqrt{\mathcal{N}}}.
\end{equation}
This expression is indeed independent of $N$ and provides a good
estimate of the standard error in the given time interval, as can
be seen from the figure. Moreover, this result implies that the
fluctuations grow with a rate which is much smaller than the one
provided by the strict upper bound $2\Gamma_0$ of
$\Gamma_++\Gamma_-$. In fact, $\Gamma_0$ scales with the square
root of $N$ which predicts a much stronger increase of the
fluctuations.

\section{Conclusions} \label{CONCLU}
It has been shown in this paper that the von Neumann dynamics of a
combined quantum system can be formulated in terms of a rather
simple piecewise deterministic process which gives rise to a
powerful and efficient Monte Carlo simulation method of the exact
non-Markovian reduced system behavior. A Markovian representation
of the dynamics was achieved through the use of a pair of product
states $\psi_{\nu}\otimes\chi_{\nu}$ in the state space of the
total system. The stochastic propagation of an ensemble of such
pairs then enables one to mimic the exact time-evolution of the
reduced system's density matrix.

The examples discussed in Sec.~\ref{SECBOSONIC} and \ref{SPINBAD}
illustrate the generality of the method: It is applicable to both
bosonic and spin environments and is not restricted to linear
dissipation or to a perturbation treatment of the
system-environment coupling. Most importantly, the method does not
require the derivation, not even the existence of a master
equation of the reduced system. At the same time, the technique
allows the direct determination of all kinds of multitime quantum
correlation functions. Although our discussion was carried out in
the interaction picture, it is obvious that the stochastic
dynamics can also be formulated in the Schr\"odinger picture, in
which case both $\psi_{\nu}$ and $\chi_{\nu}$ follow, in general,
a non-trivial deterministic evolution. Furthermore, it should be
clear that, instead of using a PDP, one can also employ a
diffusion process (Brownian motion) to construct an unraveling of
the von Neumann equation.

The stochastic technique was formulated here as a method of
simulating the dynamics of open systems in real time. A potential
extension of the method is to re-formulate the dynamics in
imaginary time \cite{CARUSO2}, in order to determine the
properties of the system in thermodynamic equilibrium. With the
total Hamiltonian $H=H_S+H_E+H_I$ in the Schr\"odinger picture the
canonical equilibrium density matrix (not normalized) is given by
$\rho(\beta) = e^{-\beta H}$, where $\beta=1/k_{\mathrm{B}}T$ is
the inverse temperature. At infinite temperature we have
$\rho(\beta=0)=I$. This suggests determining the equilibrium
density at finite temperature by solving the evolution equation
\begin{equation} \label{EVOL-IMAG}
 \frac{d}{ds} \rho(s) = -\frac{1}{2} \left\{ H , \rho(s) \right\}
\end{equation}
over the interval from $s=0$ to $s=\beta$. This imaginary-time
dynamics can again be represented in terms of a stochastic process
for a pair of product states
$|\Phi_{\nu}(s)\rangle=\psi_{\nu}(s)\otimes\chi_{\nu}(s)$. An
appropriate system of stochastic differential equations in the
Schr\"odinger picture is given by
\begin{eqnarray}
 d\psi_{\nu}
 &=& -\frac{1}{2} H_S \psi_{\nu} ds  \nonumber \\
 &~& + \sum_{\alpha}
 \left( -\frac{1}{2} L_{\alpha\nu} A_{\alpha}-I\right)
 \psi_{\nu} dN_{\alpha\nu}, \label{STOCH1-IMAG} \\
 d\chi_{\nu} &=& \left( -\frac{1}{2} H_B
 + \Gamma_{\nu} \right) \chi_{\nu} ds \nonumber \\
 &~& + \sum_{\alpha} \left( M_{\alpha\nu} B_{\alpha}-I\right)
 \chi_{\nu} dN_{\alpha\nu}. \label{STOCH2-IMAG}
\end{eqnarray}
Performing a calculation analogous to the one of Sec.~\ref{PDP} it
is easy to verify that the expectation value
$\rho(s)={\mathrm{E}}(|\Phi_1(s)\rangle\langle\Phi_2(s)|)$
satisfies the evolution equation (\ref{EVOL-IMAG}). The
$dN_{\alpha\nu}(s)$ are again independent Poisson increments
satisfying ${\mathrm{E}}(dN_{\alpha\nu}(s))=\Gamma_{\alpha\nu}ds$,
and the relations (\ref{DEFGNU}) and (\ref{DEFGAN}) remain valid.

An important restriction of the Monte Carlo technique is provided
by the behavior of the statistical fluctuations. The
considerations of Sec.~\ref{FLUCTUATIONS} as well as the example
discussed in Sec.~\ref{STOCHSIMSPIN} reveal that the method as
formulated in Sec.~\ref{PDP} is feasible, in general, only for
short and intermediate time scales. For large times statistical
errors may grow exponentially fast, ruling out the estimation of
statistical quantities with reasonable effort. However, this
conclusion rests on the assumption that the stochastic states 
$|\Phi_{\nu}(t)\rangle$ are tensor products of certain system and 
environment states. This
leads to a further potential generalization of the method, namely
to introduce a class of stochastic states with a more complicated
structure, the aim being a more efficient representation of
$\rho(t)$ as the expectation value over the corresponding random
process.

Since the interaction generally creates correlations between the
states of system and environment it could be advantageous, e.~g.,
to use a class of {\em{entangled}} stochastic states. The spin
bath model studied in Sec.~\ref{SPINBAD} leads to a trivial
example: The class of entangled states defined by
($\alpha$ and $\beta$ are complex amplitudes)
\begin{equation}
 \alpha |+\rangle \otimes |j,m\rangle
 + \beta |-\rangle \otimes |j,m+1\rangle
\end{equation}
yields an extremely efficient stochastic representation of the
dynamics: As a consequence of the conservation of the 3-component
of the total spin angular momentum, the subspaces spanned by the
states $|+\rangle \otimes |j,m\rangle$ and $|-\rangle \otimes
|j,m+1\rangle$ are invariant under the time-evolution and, thus,
the dynamics may be expressed entirely though an appropriate
(deterministic) time-dependence of the amplitudes $\alpha$ and
$\beta$. Therefore, only the initial state is a random quantity
and the statistical errors are constant in time.

In a further possible extension of the method one could employ a
stochastic evolution of {\textit{mixed}} states instead of pure states.
As an example we introduce a stochastic matrix
\begin{equation}
 R(t) = |\psi_1(t)\rangle\langle\psi_2(t)| \otimes R_E(t),
\end{equation}
where the $\psi_{\nu}(t)$ are random states of the open system and
$R_E(t)$ is a random operator in ${\mathcal{H}}_E$, and try again to
find stochastic evolution equations such that the exact von
Neumann dynamics is recovered by means of the expectation value
$\rho(t)={\mathrm{E}}(R(t))$. This is indeed possible if we use
the stochastic differential equations (\ref{STOCH1}) for the
$\psi_{\nu}(t)$ and if we replace (\ref{STOCH2}) by the following
stochastic differential equation for the random operator $R_E(t)$,
\begin{eqnarray}
 dR_E &=& \Gamma R_E dt
         +\sum_{\alpha} \left( M_{\alpha 1} B_{\alpha} - I \right)
         R_E dN_{\alpha 1} \nonumber \\
      &~& +\sum_{\alpha} R_E \left( M_{\alpha 2}
         B^{\dagger}_{\alpha} - I \right) dN_{\alpha 2},
\end{eqnarray}
where $\Gamma=\Gamma_1+\Gamma_2=\sum_{\alpha\nu}
\Gamma_{\alpha\nu}$ is the total jump rate. The further
development of the stochastic technique proposed in this paper
should include a systematic investigation of the potentialities of
the extensions indicated above.

\begin{acknowledgments}
The author would like to thank F.~Petruccione for
helpful discussions and comments.
\end{acknowledgments}

\end{document}